\documentclass[10pt,journal]{IEEEtran}
\PassOptionsToPackage{colorlinks=true,linkcolor=blue,citecolor=blue,urlcolor=blue}{hyperref}
\PassOptionsToPackage{pdftex}{graphicx}
\makeatletter
\AtBeginDocument{%
  \DeclareGraphicsExtensions{.pdf,.png,.jpg}%
}
\makeatother

\usepackage{ifpdf}
\usepackage{caption}
\usepackage{amssymb}
\usepackage{pifont}
  \usepackage{cite}

\ifCLASSINFOpdf
  \usepackage[pdftex]{graphicx}
  
\else
  
\fi
\usepackage[T2A]{fontenc}          
\usepackage[utf8]{inputenc}        
\usepackage[english,russian]{babel}

\usepackage{graphicx}
\DeclareGraphicsExtensions{.pdf,.png,.jpg}

\usepackage{amsmath,amssymb,amsthm}
\usepackage{microtype}

\usepackage[colorlinks=true,linkcolor=blue,citecolor=blue,urlcolor=blue]{hyperref}

\usepackage{stfloats}
\usepackage{amsmath}
\usepackage{amssymb}
\usepackage{acronym}
\usepackage{array}
\usepackage{mdwmath}
\usepackage{mdwtab}
\usepackage{xcolor}
\usepackage{multicol}
\usepackage{tabularx}
\usepackage{longtable}
\usepackage{multirow}
\usepackage{subfig}
\usepackage{algorithm}
\usepackage{url}
\usepackage{etoolbox}
\usepackage{stackengine}

\usepackage{hhline}
\usepackage{booktabs}
\usepackage{todonotes}
\usepackage{graphicx}
\usepackage{booktabs}

\usepackage{stackengine}
\usepackage{amssymb}

\usepackage{url}
\usepackage{listings}
\usepackage{xcolor}

\usepackage{graphicx}
\usepackage{adjustbox}
\usepackage{booktabs}

\definecolor{codegray}{gray}{0.95}
\definecolor{commentgray}{rgb}{0.3,0.3,0.3}
\definecolor{keywordblue}{rgb}{0,0,0.6}
\definecolor{stringred}{rgb}{0.58,0,0.01}

\lstdefinelanguage{CSharp}{
  morekeywords={public, private, void, float, int, new, class, using, return, if, for, in, Debug},
  sensitive=true,
  morecomment=[l]{//},
  morestring=[b]",
}

\begin{document}
\title{Meta-Guardian: An Early Evaluation of an On-device Application to Mitigate Psychography Data Leakage in Immersive Technologies}
\author{Keshav Sood,~Sanjay Selvaraj,~and Youyang Qu
\IEEEcompsocitemizethanks{\IEEEcompsocthanksitem Keshav Sood and Sanjay Selvaraj is with Deakin University, Australia. E-mail: keshav.sood@deakin.edu.au, sanjay.selvaraj@deakin.edu.au\\ Youyang Qu is with Data61, Commonwealth Scientific and Industrial Research Organization (CSIRO), Sydney, Australia. E-mail: youyang.qu@data61.csiro.au}}

\markboth{IEEE}%
{Shell \MakeLowercase{\textit{et al.}}: Bare Advanced Demo of IEEEtran.cls for IEEE Computer Society Journals}
\IEEEtitleabstractindextext{%
\begin{abstract}
The use of Immersive Technologies has shown its potential to revolutionize many sectors such as health, entertainment, education, and industrial sectors. Immersive technologies such as Virtual Reality (VR), Augmented reality (AR), and Mixed Reality (MR) have redefined user interaction through real-time biometric and behavioral tracking. Although Immersive Technologies (XR) essentially need the collection of the biometric data which acts as a baseline to create immersive experience, however, this ongoing feedback information (includes biometrics) pose critical privacy concerns due to the sensitive nature of the data collected. A comprehensive review of recent literature explored the technical dimensions of related problem, however, they largely overlook the challenge particularly the intricacies of real-time biometric data filtering within head-mounted display system. Motivated from this, in this work, we propose a novel privacy-preserving system architecture that identifies and filters biometric signals (within the VR headset) in real-time before transmission or storage. Implemented as a modular Unity Software-development Kit (SDK) compatible with major immersive platforms, our solution (named Meta-Guardian) employs machine learning models for signal classification and a filtering mechanism to block sensitive data. This framework aims to enable developers to embed privacy-by-design principles into immersive experiences on various headsets and applications.
\end{abstract}

\begin{IEEEkeywords}
User privacy, immersive technologies, virtual reality, data privacy, software security, cyber security.
\end{IEEEkeywords}}

\maketitle

\IEEEdisplaynontitleabstractindextext

\IEEEpeerreviewmaketitle

\ifCLASSOPTIONcompsoc
\IEEEraisesectionheading{\section{Introduction and Background}\label{sec:introduction}}
\else
\section{Introduction}
\label{sec:introduction}
\fi

%
%
%
%
\IEEEPARstart{I}{mmersive Technologies}, such as Virtual Reality (VR) and Augmented Reality (AR), and Mixed Reality (MR) enable users to experience and interact with computer-generated environments in a manner that simulates physical presence and immersion~\cite{10384639}. Immersive Technologies (XR) sometimes interchangeably called ``metaverse" is a dynamic and interconnected digital realm, holds immense potential to reshape the way we live, work, and interact; it blurs the boundaries between physical and (semi-)virtual spaces. The core concept of VR centers is on the creation of synthetic spaces that can be explored and manipulated, typically through the use of head-mounted displays (HMDs), motion tracking, and various input devices. These systems are designed to provide a multi-sensory experience, often incorporating visual, auditory, and increasingly haptic feedback to enhance the sense of immersion~\cite{10938253}. 
\par 
The adoption of XR has accelerated in recent years, driven by advances in hardware capabilities, reductions in cost, and the proliferation of content across entertainment, education, healthcare, and industrial training~\cite{lee2025conceptual, heller2020reimagining}. A defining feature of XR systems is their capacity to collect and process a wide array of user data, including behavioral and biometric signals~\cite{bernadelli2021dynamic, chen2022query2set}. The integration of biometric data collection within XR environments has become increasingly prevalent, driven by the pursuit of more immersive and personalized user experiences. Biometric signals such as heart rate, facial expressions, gaze direction, and head movement are now routinely captured to enable adaptive content, enhance interaction fidelity, and support continuous authentication mechanisms~\cite{9880528, shi2021face, 8919581}. These data streams, while offering significant benefits for usability and security, however, introduce a new dimension of privacy and security challenges that are distinct from those encountered in traditional digital systems~\cite{sood2025framework, senthuran2025balancing}. \par 
Note that user safety and privacy is vital in XR space and inadequate privacy measures can lead to severe online safety risks\footnote{https://www.afp.gov.au/news-centre/media-release/holiday-season-warning-extremists-infiltrating-online-and-gaming, accessed on 02/208/2025}; for example, the absence of enforceable community guidelines or mechanisms that promote user privacy may lead to invasions of privacy since the biomertic data collection by these systems are without user consent. The sensitivity of biometric data is underscored by its inherent link to an individual’s physiological and behavioral characteristics, making it uniquely identifying and difficult to revoke or alter in the event of a breach~\cite{agarwal2024biometrics}. Unlike passwords or tokens, biometric traits such as facial geometry or cardiac rhythms are immutable, and their exposure can have long lasting consequences for user privacy~\cite{gomez2024scoping}.  Also, as a user we believe that we are participating in private or sensitive settings, only for another user to record their behaviors and share it without consent\footnote{https://www.weforum.org/stories/2024/08/metaverse-interoperability-regulation/, accessed on 02/08/2025}. So, it becomes imperative to address such concerns to ensure a secure and inclusive metaverse for all participants. \par
We~\cite{sood2025framework} reported that the use of photorealistic avatars and the embedding of biometric features in extended reality (XR) systems further amplify these risks, as attackers may exploit such data to impersonate users or infer sensitive health information~\cite{sood2025framework} leads to cognitive manipulation of users and a violation of human rights\footnote{https://humanrights.gov.au/our-work/technology-and-human-rights/publications/protecting-cognition-background-paper, accessed on 02/08/2025}. The date collected through users’ engagement with XR systems may be misused for malicious purposes. \textit{``This may include physical harm towards individuals identified as belonging to marginalised or minority groups or identifying and doxing victim-survivors of family, domestic and sexual violence who may be seeking refuge in undisclosed locations"}\footnote{https://www.esafety.gov.au/sites/default/files/2025-05/Immersive-technologies-position-statement-May-2025.pdf, accessed on 01/08/2025}. Our early work in~\cite{sood2025framework} serves as a catalyst for users, governments across the world, stakeholders, promoting dialogue and action in navigating the complexities of privacy and safety in this blended and blur XR world.\par
Recent advances in VR technology have led to the adoption of continuous authentication strategies, where user identity and intent are verified in real time through ongoing analysis of biometric and behavioral cues~\cite{10159439, luo2021fa}. This approach enhances security by dynamically adapting to user behavior and detecting anomalies that may signal unauthorized access. However, it also necessitates the persistent collection and processing of sensitive data, raising concerns about potential leaks, unauthorized access, and misuse both within and beyond the VR device. The risk landscape is further complicated by the integration of VR systems with broader metaverse infrastructures, which often involve interconnected platforms, cloud services, and third-party applications~\cite{agarwal2024biometrics}. These complex ecosystems introduce additional attack vectors, including data interception during transmission, unauthorized aggregation of biometric profiles, exploitation of system vulnerabilities\footnote{https://arstechnica.com/gaming/2024/02/meta-will-start-collecting-anonymized-data-about-quest-headset-usage/, accessed on 03 June 2025}. \par 
To address these challenges, we emphasize that several risk factors are needed to be considered during the design, development, and deployment of XR products. We emphasize that users have difficulties in XR environments to capture evidence of harm, therefore it is difficult for regulators, platforms and law enforcement agencies to detect harm. This is particularly due to the fleeting nature of immersive interactions and the absence of URLs to pinpoint locations of harm. Therefore, it requires some level of targeted and specific design approaches to address the challenges posed by XR technologies. \par 
In our previous works~\cite{sood2025framework, senthuran2025balancing}, we have proposed frameworks that prioritize user privacy by ensuring that biometric data remains confined to the local device, thereby minimizing the risk of external exposure. Sood et al.~\cite{sood2025framework} have stated that a two-stage methodology for biometric signal extraction and secure processing can provide a foundation for privacy-preserving innovation in immersive technologies. Such approaches are complemented by the adoption of zero-trust security architectures, which emphasize user control over data, identity, and experience, and implement proactive measures to safeguard against both internal and external threats. The usability of VR systems is closely intertwined with the effectiveness of their privacy and security mechanisms. While biometric authentication can streamline access and enhance convenience, it must be balanced against the potential for unauthorized access and data breaches. The implementation of privacy-preserving technologies, such as on-device encryption, is essential to maintaining user trust and supporting the widespread adoption of immersive applications. \par 
\textbf{\textit{Novelty:}} While prior research has proposed frameworks~\cite{sood2025framework, senthuran2025balancing, heller2020reimagining, yang2023prediction, wang2023identifying, o2023privacy, wu2023privacy, zhang2023facereader} aimed at addressing biometric privacy concerns in XR environments often focusing on post-processing anonymization or policy based access control these approaches may fall short in scenarios demanding real-time protection or user-level customization, particularly on-device. To bridge this gap, in this work we introduce a novel and proactive framework that operates at the point of data generation (on-device), specifically targeting the real-time detection and suppression of sensitive biometric signals related to eye-tracking and facial features.\par
Our contributions in this paper are below. 
\begin{enumerate}
    \item We pointed out that the biometric data leakage must be blocked within the VR device only and should not be allowed to be collected, stored, and processed outside the device without user consent (which is currently being ignored). Unlike reactive methods that rely on downstream anonymization or external oversight, we proposed a Software Development Kit (SDK) that functions as an embedded privacy filter within the VR pipeline (device itself), intercepting biometric data before it can be logged, stored, or transmitted out of the device. 
    \item Our proposed SDK continuously monitors XR data streams to identify biometric patterns such as gaze direction, pupil movement, blink frequency, and facial expressions. By integrating biometric-aware filtering directly into the runtime environment, we reduce the attack surface for potential misuse and provide a technical safeguard against unauthorized profiling or surveillance.
    \item Furthermore, the framework is designed with strong alignment to contemporary privacy regulations and ethical design principles. It enforces data minimization by ensuring that only essential, nonbiometric data is retained; adheres to purpose limitation by preventing biometric data from being repurposed without consent; and enables user-centric control, allowing individuals to define and update their privacy preferences dynamically.
    \end{enumerate}
\par 
\textbf{\textit{Benefits:-}} Overall, in this work we have taken an initiative which will encourage technology companies to take proactive steps to secure and maintain user safety and privacy. It will further push them to invest in risk mitigation at the forefront – and throughout all stages – of product design, development, and deployment. Embedding safer design practices (such as the one we propose) will minimize the risk of harm from XR technologies. Safety by Design (although voluntary) principles (the proposed design/framework) will ensure who is held accountable for (service provider responsibility), it will provide user more autonomy and empowerment, and bring transparency and accountability in the rollover and usage of the technology~\cite{o2023privacy}. Finally, industries can use this as a way to support compliance with regulatory requirements\footnote{https://www.esafety.gov.au/industry/tech-trends-and-challenges/immersive-tech, accessed on 01/08/2025}.


\section{Related Work}
In Metaverse, \textit{``the sensory information and actuator-related information are exchanged between virtual and real worlds via IEEE 2888.1 and IEEE 2888.2 standards, respectively. Besides, the definition, synchronization, and mission control data are defined by the IEEE 2888.3 standard for digital things (i.e., virtual objects)"}~\cite{9880528}. No doubt there are standards for metaverse and the other enabling technologies as well, however, most of the literature we have studied  focused on some typical issues such as Identity Theft\footnote{https://threatpost.com/nft-investors-lose-1-7m-in-opensea-phishing-attack/178558/, accessed on 06/11/2023}, Impersonation Attack, Avatar Authentication Issue,  Trusted and Interoperable Authentication,  Unauthorized Data Access, Misuse of User/Avatar Data, etc.~\cite{9880528, jiang2021reliable, han2022dynamic, lee2021creators, huynh2023artificial, falchuk2018social, ning2023survey}. To the best of our literature review, we note that very little has been explored to mitigate the issue of Biometrically inferred data~\cite{9880528, ning2023survey, huynh2023artificial, lee2021creators}. Previous works related to on-device data detection and suppression of biometric signals represent a proactive approach to privacy protection, particularly relevant in immersive environments where real-time data acquisition is integral to user experience. The core idea is to intercept and process biometric data at the point of collection, ensuring that sensitive information is either filtered or obfuscated before it leaves the user’s device. This strategy directly addresses the risk of data leakage and unauthorized access, which are not mitigated by traditional biometric authentication methods alone \cite{rai2023survey},\cite{topcu2016practical}. \par
The technical implementation of on-device suppression often relies on advanced signal processing techniques. For instance, the use of Gabor filters, Laplace filters, and Volter filters has been shown to enhance the quality of biometric images, such as fingerprints, by transforming them in both frequency and spatial domains. Adjusting the parameters of these filters can significantly improve the visibility of structural features while also enabling selective suppression of sensitive attributes~\cite{nazarkevych2019detection, dick2021balancing}. However, conventional filtering methods may not be sufficient for robust privacy protection, as adversaries can potentially reconstruct original biometric features from processed data, especially if the transformation is invertible or if auxiliary information is available. \par 
To counteract such vulnerabilities, privacy-enhancing technologies must be integrated at the acquisition stage. Differential privacy and homomorphic encryption are promising approaches that can be applied on-device. Differential privacy introduces controlled noise to the data, making it statistically improbable to infer individual-specific information, even if the processed data is accessed by unauthorized parties. Homomorphic encryption, on the other hand, allows computations to be performed on encrypted data, ensuring that raw biometric signals remain inaccessible throughout the processing pipeline. These methods, when implemented on-device, guarantee that sensitive biometric information is never exposed in plaintext, reducing the attack surface for potential adversaries. \par 
However, the challenge of real-time detection and suppression is further complicated by the diversity of biometric modalities used in immersive technologies. EEG, ECG, eye movement, and facial expressions each present unique signal characteristics and privacy risks. For example, multimodal data fusion, such as combining EEG and ECG, can inadvertently increase the risk of sensitive information leakage if not properly managed, as the fusion process may amplify the discriminative power of the combined features \cite{yang2023prediction},\cite{drozdowski2020multi}. Therefore, on-device suppression mechanisms must be modality-aware, capable of identifying and filtering out sensitive features specific to each signal type before any data leaves the device. \par 
Recent research has also highlighted the importance of adversarial perturbations as a means of suppressing identifiable attributes in biometric data. A comprehensive review of recent literature \cite{rai2023survey,topcu2016practical,nazarkevych2019detection,yang2023prediction,drozdowski2020multi} indicates that, although several studies have explored the technical dimensions of related problems, they largely overlook the fundamental challenges associated with immersive technologies particularly the intricacies of real-time biometric data filtering within head-mounted display systems~\cite{sood2025framework}. This gap highlights the need for targeted research that bridges system-level optimization with privacy-preserving mechanisms in immersive environments.

\section{{Meta-Guardian: Methodology and Design}}
\begin{figure}
    \centering
    \includegraphics[width=1\linewidth]{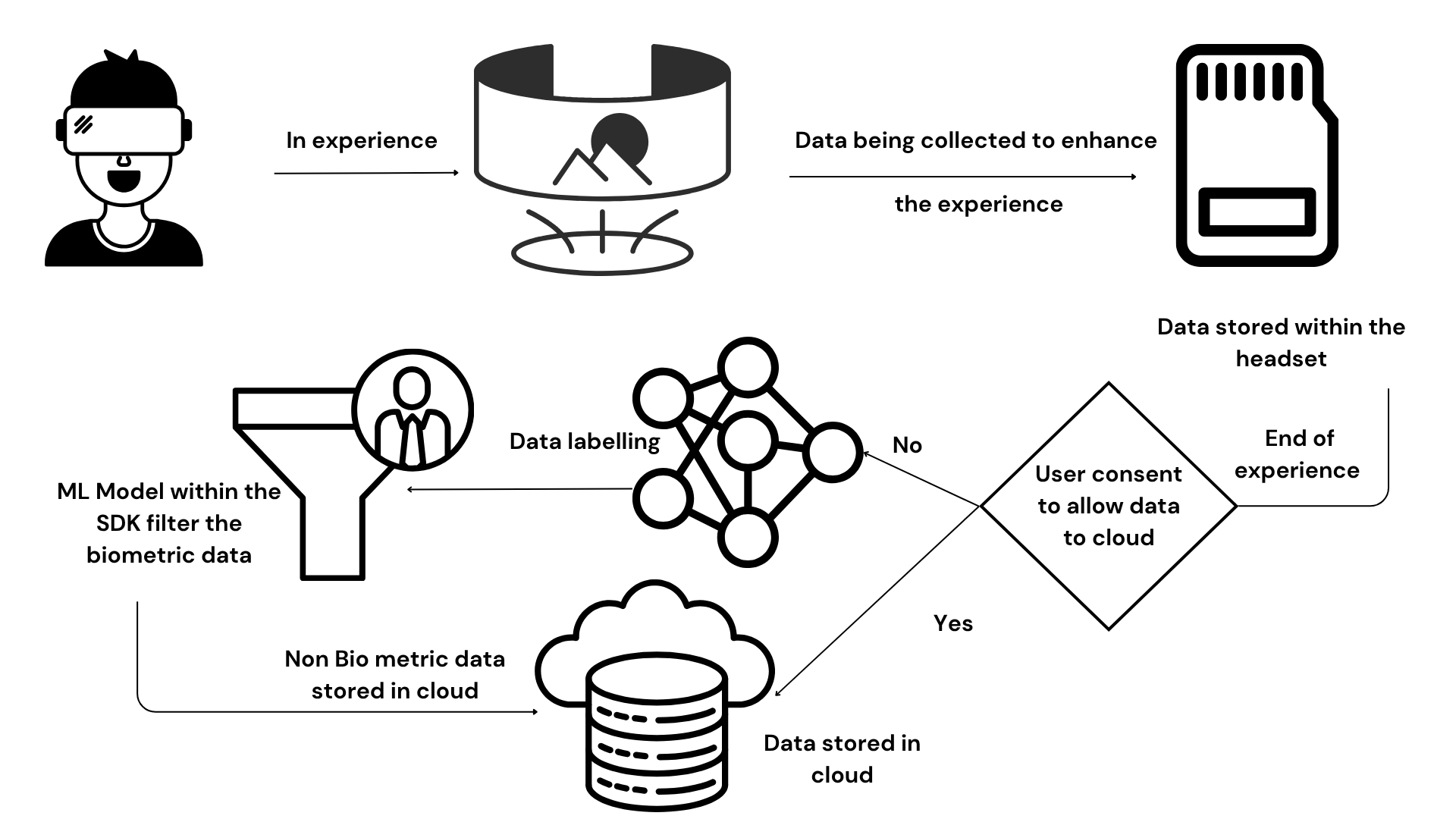}
    \caption{Framework Architecture}
    \label{fig:1}
\end{figure}
The architecture of our proposed framework, shown in Fig.~\ref{fig:1} is designed within the VR device to ensure the real-time identification and suppression of facial and ocular biometric signals in extended reality environments, with a focus on compliance with privacy regulations and minimal disruption to application performance. This system operates through three sequential stages: 1) Signal Collection, 2) Biometric Filtering and  Enforcement, and 3) SDK design. \par 
\textbf{\textit{{Signal Collection:}}} Our system leverages the Meta Quest Pro headset, which includes advanced inward-facing sensors capable of capturing high-fidelity facial expressions and eye movement data at high temporal resolution. The data collected is both spatially detailed and temporally synchronized, forming a rich source for identifying biometric content. 1) \textit{Facial Expression Data:} The Meta Quest Pro provides expression weights corresponding to fine-grained muscle activations. These include features such as: Eyebrow movements, Cheek activity, Jaw dynamics, Lip movements. 2) \textit{{Eye Tracking Data:}} The system also captures real-time metrics related to ocular behavior such as: Gaze vector, blink rate, and eye openness.

\subsection{Biometric Filtering and Privacy Enforcement }
In the final stage, signal vectors identified as containing biometric content are intercepted at runtime. Depending on the configuration set by the application developer or end-user preferences, these signals can be 1)~\textit{Suppressed:} means the biometric component is removed from the processing pipeline entirely, or~\textit {Passed Through:} i.e., if non-sensitive or explicitly consented by the user. This selective filtering occurs prior to any storage, visualization, or transmission, ensuring that sensitive biometric data never leaves the device unprotected. The framework is implemented as a lightweight SDK compatible with existing Unity-based XR workflows. Developers can integrate the privacy module with minimal code-level changes, and it offers customizable privacy presets aligned with data minimization, purpose limitation, and user consent enforcement as per GDPR and similar standards.\par 
\subsection{Development of the proposed SDK: Implementation Guide:} To support real-time biometric filtering within XR environments, we developed a modular software development kit (SDK) implemented in Unity\footnote{https://docs.unity3d.com/Manual/building-and-publishing.html, accessed on 03 July 2025}. The SDK is designed for extensibility and performance, offering seamless integration with the Unity XR pipeline and third-party SDKs such as Oculus Integration (Meta)\footnote{https://liblab.com/blog/how-to-build-an-sdk, accessed on 03 July 2025}. Our SDK follows a component-based, interface-driven design that promotes decoupling and ease of integration. Each stage in the pipeline is encapsulated within well-defined interfaces, allowing developers to swap or extend modules without altering core logic\footnote{https://liblab.com/blog/sdk-vs-api, accessedon 03 July 2025}. The SDK follows an interface-driven architecture (modular architecture), where each processing stage is encapsulated in distinct modules. This structure supports plug-and-play customization and simplifies adaptation to various XR platforms and sensor configurations. The core modules include: 

\textit{1. Signal Collector}: Provides device-agnostic access to raw input signals such as facial muscle activations, gaze vectors, and physiological metrics. The abstraction layer enables easy integration of multiple hardware vendors without modifying downstream components. 

\textit{2. Feature Processor}: Extracts high-level biometric features from raw data in real time. These features may include blink rate, gaze fixation entropy, expression intensity scores, and muscle activation symmetry. The module is optimized for frame-aligned execution in Unity’s update cycle. 

\textit{3. Biometric Classifier}: Applies lightweight machine learning models to identify whether a given feature vector is likely to reveal biometric information. Models are trained offline using labeled datasets and exported in ONNX format. Runtime inference is performed using Unity’s Barracuda inference engine, ensuring compatibility with standalone XR headsets. 

\textit{4. Biometric Filter}: Implements the enforcement logic, allowing biometric data to be suppressed, or retained based on classification outcomes and user-defined privacy selection. 

\textit{5. Consent Manager}: This is the critical component ensuring that no biometric data leaves the device without informed user approval. Before any signal is transmitted to a remote server or cloud service, the Consent Manager triggers a contextual user prompt. The user may allow, deny, or selectively permit certain features to be shared. 
In our work, the SDK currently supports on-device workflows. In the on-device configuration, all signal processing, classification, and filtering occur locally, ensuring that raw data remain confined to the headset. This mode is recommended for privacy-sensitive applications.  \par

\subsection{On VR Device and Platform Rationale}

\textit{1. The Meta Quest Pro} was chosen for this research due to its advanced built-in sensor suite, which includes five inward-facing infrared cameras for facial expression capture and two eye-facing cameras for accurate real-time gaze tracking. This integrated hardware allows the headset to provide detailed expression weight data and eye movement vectors without requiring external tracking systems, making it ideal for capturing rich biometric signals in a naturalistic and untethered setting. Its lightweight, ergonomic design improves user comfort during extended sessions an essential consideration for immersive experimental tasks. Furthermore, Quest Pro supports high-frequency sampling of eye tracking (up to 90 Hz) and facial expression data (\~30 Hz), offering a level of temporal resolution well suited for behavioral studies and biometric modeling.\par
\textit{2. The Unity game engine} was selected as the development platform due to its robust XR development pipeline, wide adoption in both academic and commercial VR projects, and seamless integration with the Meta XR SDK. Unity provides direct access to the headset's facial and eye tracking data streams and supports rapid prototyping of interactive environments within a modular architecture. Its real-time rendering capabilities and plugin ecosystem further enabled flexible control over stimulus presentation, data logging, and system calibration, making it an ideal environment for building immersive, headset-compatible experimental applications.

\section{Threat Model }
To establish the scope and evaluate the effectiveness of our proposed biometric privacy-preserving framework, we define a formal threat model that outlines the adversary's objectives, capabilities, and attack surfaces within immersive XR systems.\par  

\textit{{1. Adversary Objectives:-}} The adversary’s primary goal is to exploit the biometric signals captured by AR/VR headsets to infer sensitive or private information about the user. This may include: a) Re-identify a user based on subtle and consistent biometric patterns (e.g., unique facial motion signatures during interaction), b) Infer emotional or cognitive states, (e.g., anger, fatigue, stress) without user awareness, using signals like brow tension, lid tightening, or cheek raising, c) Predict user reactions to stimuli, such as determining which parts of an emotional video elicited surprise (via inner brow raiser) or disgust (via upper lip raiser), d) Profile behavioral tendencies, such as hesitation patterns, eye-gaze consistency, or muscle fatigue across sessions, to develop psychography models, and e) Extract speech intent or content using facial motion vectors (e.g., jaw drop + tongue motion), even when the microphone is disabled. 

\textit{{2. Adversary Capabilities:-}} We assume a software-level adversary with considerable access, but not full system control. They may include: a) Third-Party XR Applications or plugins operating within the Unity engine or headset runtime, these apps can access sensor APIs (e.g., \verb|OculusAvatar2FaceExpression|, \verb|SRanipal|) and may collect expression weights, gaze vectors, or skeletal movement data every frame. For example: A meditation app records 10-second intervals of \verb|JawDrop|, \verb|LidTightener|, and \verb|CheekRaiser| signals. A third-party SDK logs this data and sends it to a cloud server for “UX analysis,” but can reconstruct when the user showed signs of emotional breakdown or stress from ambient stimuli. b) Machine Learning Inference Attacks: Adversaries may use trained models to correlate recorded biometric features with: demographic traits (e.g., age via blink rates), emotional reactivity (e.g., affective valence via cheek and brow activations), personality dimensions (e.g., introversion via gaze avoidance), etc., for example using datasets collected during gameplay, an adversary builds a classifier that maps specific sequences of facial weights (e.g., \verb|LipPucker| + \verb|UpperLipRaiser| + \verb|NoseWrinkler|) to negative sentiment, allowing targeted content manipulation, c) Network Eavesdropping: if raw telemetry or analytics data (e.g., user session logs) are transmitted unencrypted or with weak access control, an external attacker may extract biometric signals from packet payloads, for exxample, a cloud log shows average gaze direction, expression states, and timestamped events. Without filtering, even “non-sensitive” logs reveal that a user looked downward and exhibited jaw motion 3.2 seconds after an emotional video clip—revealing unconscious reactions.\par 

\textit{{3. Adversary Positioning:-}} we consider the following adversary classes: a) Internal Adversaries: XR app developers, SDK vendors, or embedded analytics scripts with runtime access, b) External Adversaries: Eavesdroppers intercepting data en route to cloud infrastructure or malicious endpoints receiving exported logs, and c) Curious Insiders: Developers or analysts within organizations that store or replay user session data without biometric filtering.\par

\textit{{System Assumptions:-}} Our framework is designed assuming the following. a) The XR engine (Unity), headset OS, and sensor firmware are trusted and not compromised, b) Biometric filtering is enforced locally within the runtime, not externally post-processing logs, c) Users consent to data collection but do not expect fine-grained behavioral traits to be inferred or stored without filtering. Our framework dose not protect against: a) Root-level exploits of the headset firmware or OS, b) Hardware tampering (e.g., custom sensor implants), c) Physical observation or social engineering outside the XR environment. \par

\textit{{Security Goals and Coverage:-}} Our system is built to counter the above adversaries by: a) Intervening in real time, using a lightweight ML model to classify expression weights (for example, Meta Quest Pro or Vive Eye Pro) every few frames, b) Suppressing signals that are identified as biometric, including \verb|JawDrop|, \verb|TongueMotion|, or gaze vectors with high reidentifiability, c) Preventing transmission or logging of sensitive time windows (e.g., peaks of emotional response during video playback), d) Providing a transparency layer, allowing developers to understand what is filtered and why, without relying solely on post hoc anonymization.
By embedding privacy enforcement into the Unity pipeline, before storage or transmission, we reduce the available attack surface for adversaries operating within the app layer.

\section{Experimental Design and Data Collection}

To evaluate the risks of biometric leakage from facial expressions in immersive XR, we designed three controlled virtual environments using Unity: 

\textit{1. Interactive Gaming Environment}: Participants performed fast-paced object interaction tasks (e.g., hitting, dodging, catching) that triggered strong motor and gaze-based responses. Metrics such as gaze velocity, eye orientation, head rotation, and jaw displacement were recorded. 

\textit{2. Emotional Video Playback Environment}: Participants viewed emotionally curated video clips designed to evoke contrasting affective states (e.g., joy vs. sadness). We focused on involuntary facial responses, including cheek raising, brow motion, lip tension, and blink rate. 

\textit{3. Ambient Spiritual Experience}: A calm, meditative environment with low stimulation and ambient audio encouraged a state of rest. This scene was intended to capture passive muscle relaxation, reduced jaw activity, and spontaneous blink frequency. \par

\begin{figure}[htbp]
\centering
\includegraphics[width=0.4\textwidth]{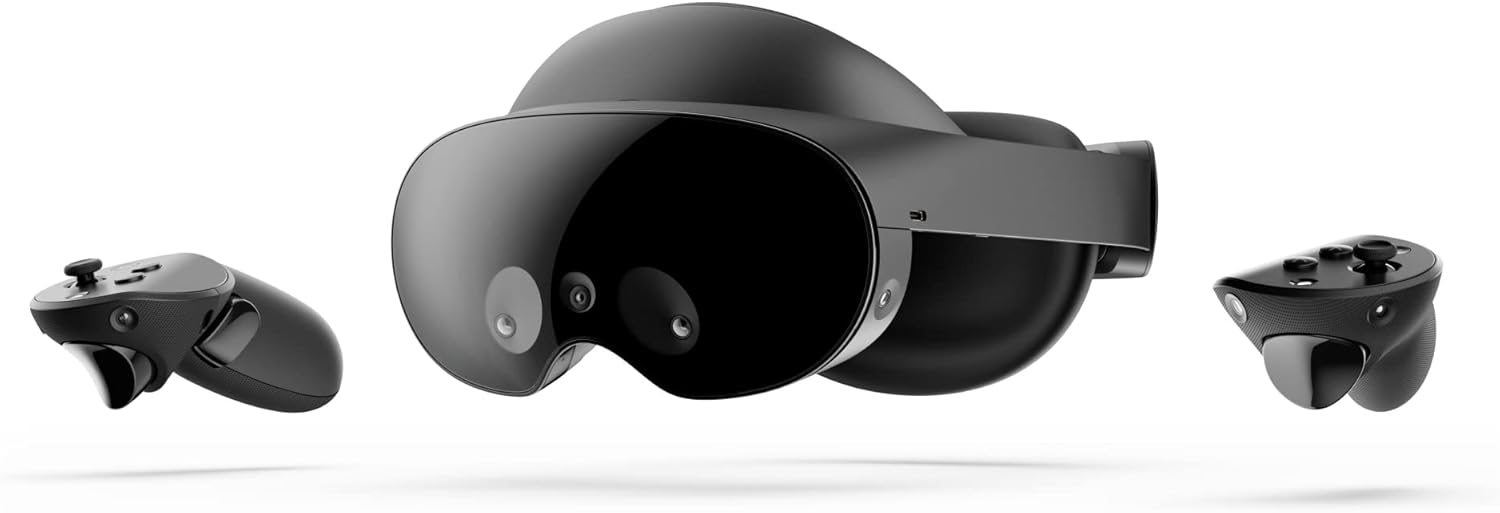}
\caption{The Meta Quest Pro headset and Touch Pro controllers, showing the inward-facing IR cameras and sleek mixed-reality form factor.}
\label{fig:quest_pro}
\end{figure}

All experiments were conducted using the Meta Quest Pro, Fig.~\ref{fig:quest_pro}, a standalone head-mounted display  equipped with five inward-facing infrared sensors specifically designed for capturing facial expressions and eye movement. The headset supports real-time acquisition of expression weight vectors, representing fine-grained muscular activations such as brow movement, cheek tension, jaw displacement, and lip deformation. Additionally, its eye-tracking system captures metrics including gaze direction, blink rate, and eye openness, sampled at a frequency of approximately $90$ Hz. Expression weights are recorded at a lower frequency, typically around $30$ Hz, sufficient for most real-time emotional state modeling tasks.\par 

The device operates wirelessly with on-device processing, and was paired with a workstation running Unity $2022.3$ LTS, integrated via the Meta XR SDK v$77.0.0$, which exposed facial blendshapes and gaze vectors through Unity’s XR Subsystems. Each headset was calibrated before each session using the built-in eye calibration tool to ensure spatial tracking accuracy. The experimental software recorded all facial and eye data locally at runtime with precise timestamps for post-session analysis. Table~\ref{tab:quest_specs} summarizes the key technical specifications of the headset used in this study. Data was sampled at 10-second intervals, resulting in around 900+ samples. The data set was then used for the training and interpretation analysis of biometric classifiers.

\begin{table}[htbp]
\captionsetup{font=small}
\caption{Meta Quest Pro Headset Specifications}
\label{tab:quest_specs}
\begin{center}
\resizebox{0.4\textwidth}{!}{%
\begin{tabular}{c|c}
\bfseries Specification & \bfseries Details \\
\hline
Display Type & QD-LCD with local dimming \\
Resolution (per eye) & 1800 × 1920 pixels \\
Refresh Rate & 72Hz, 80Hz, 90Hz (adaptive) \\
Field of View & Approx. 106° H, 96° V \\
Processor & Snapdragon XR2+ Gen 1 \\
RAM & 12 GB \\
Storage & 256 GB \\
Eye Tracking & Yes (2 IR cameras/eye) \\
Face Tracking & Yes (5 IR cameras) \\
Gaze Sampling Rate & ~90 Hz \\
Expression Sampling Rate & ~30 Hz \\
Passthrough & Full-color stereoscopic \\
Battery Life & 1.5–2.5 hours \\
Weight & ~722 g (with strap) \\
Connectivity & Wi-Fi 6E, Bluetooth 5.2 \\
Software Stack & Unity 2022.3, Meta XR SDK v77\\
\end{tabular}%
}
\end{center}
\end{table}

\subsection{Classifier Architecture and Training}
To enable privacy-aware, on-device classification of user affective and behavioral states, we developed a lightweight deep learning classifier designed specifically for deployment on the Meta Quest Pro headset. The system performs real-time inference on biometric features derived from facial expression weights and gaze tracking data, offering robust predictions without requiring cloud connectivity or backend computation.

\subsection{Architecture and Model Design}

The classification model is implemented as a multilayer perceptron (MLP), constructed using PyTorch and optimized for mobile execution. Each model input is a 14-dimensional feature vector, extracted at 10-second intervals from the Meta Quest Pro's facial expression APIs and eye tracking sensors. Features include normalized intensity values for select facial action units (e.g., \verb|CheekRaiserL|, \verb|BrowLowererR|, \verb|LipCornerPullerL|) and real-time gaze direction metrics (e.g., \verb|EyesLookRightL|, \verb|EyesLookDownR|), chosen for their high information gain in emotional and behavioral inference tasks. The network architecture consists of Input Layer: 14 nodes (one per normalized feature), Hidden Layer: 64 fully connected units with ReLU activation to introduce non-linearity and learn complex interdependencies between features, Dropout Layer: with dropout probability \textbf{p = 0.3}, to mitigate overfitting and enhance generalization on small-scale datasets. Output Layer: multi-node Softmax classifier corresponding to a multi-class target space (e.g., \verb|Neutral|, \verb|Engaged|, \verb|Stressed|, \verb|Relaxed|). Training was conducted using the Adam optimizer with a learning rate of 0.001, batch size of 32, and 100 epochs. These hyperparameters were selected based on empirical tuning across multiple cross-validation folds.

\subsection{Data Collection and Labeling Strategy}

The biometric dataset comprises 930 labeled samples collected from participants interacting in three distinct VR environments. 1) Interactive: Task-based scenes involving active gaze, gesture, and attention redirection. 2) Emotionally Evocative: Simulated social scenarios designed to trigger affective facial responses (e.g., smiles, frowns, surprise). 3) Passive/Ambient: Meditative or low-stimulus environments intended to induce calm or neutral expressions. Each biometric vector is associated with a predefined user state label, assigned based on the specific activity performed and post-session observation of participant reactions. For example, In the interactive environment, sustained gaze and upper-lip tension might be labeled \verb|Engaged| and in the ambient environment, low cheek activation and minimal brow movement are tagged as \verb|Relaxed|. The complete dataset was normalized using z-score scaling, ensuring all features have zero mean and unit variance, Stratified into training and testing sets using a 70/30 split to preserve class balance, and evaluated using 5-fold cross-validation, ensuring model robustness across environment conditions and user variability.

\subsection{Deployment Pipeline: From PyTorch to Unity}

\begin{table}[htbp]
\captionsetup{font=small}
\caption{Explanation of Unity Sentis C\# Script Components}
\label{tab:sentis_explanation}
\begin{center}
\begin{tabular}{c|c}
\bfseries Code Element & \bfseries Purpose \\
\hline
\texttt{onnxModelAsset} & ONNX model reference in Unity \\
\texttt{ModelLoader.Load()} & Loads the model at runtime \\
\texttt{CreateWorker()} & Initializes Sentis inference worker \\
\texttt{featureVector} & Holds input from facial/gaze data \\
\texttt{Tensor(1, 14, ...)} & Converts input to tensor format \\
\texttt{worker.Execute()} & Runs inference on current input \\
\texttt{PeekOutput()} & Retrieves prediction result \\
\texttt{ArgMax()} & Gets predicted class index \\
\texttt{Dispose()} & Frees memory (prevents leaks) \\
\end{tabular}
\end{center}
\end{table}

\begin{figure}
    \centering
\subfloat[]{\includegraphics[width=3.32in, height=1.6in]{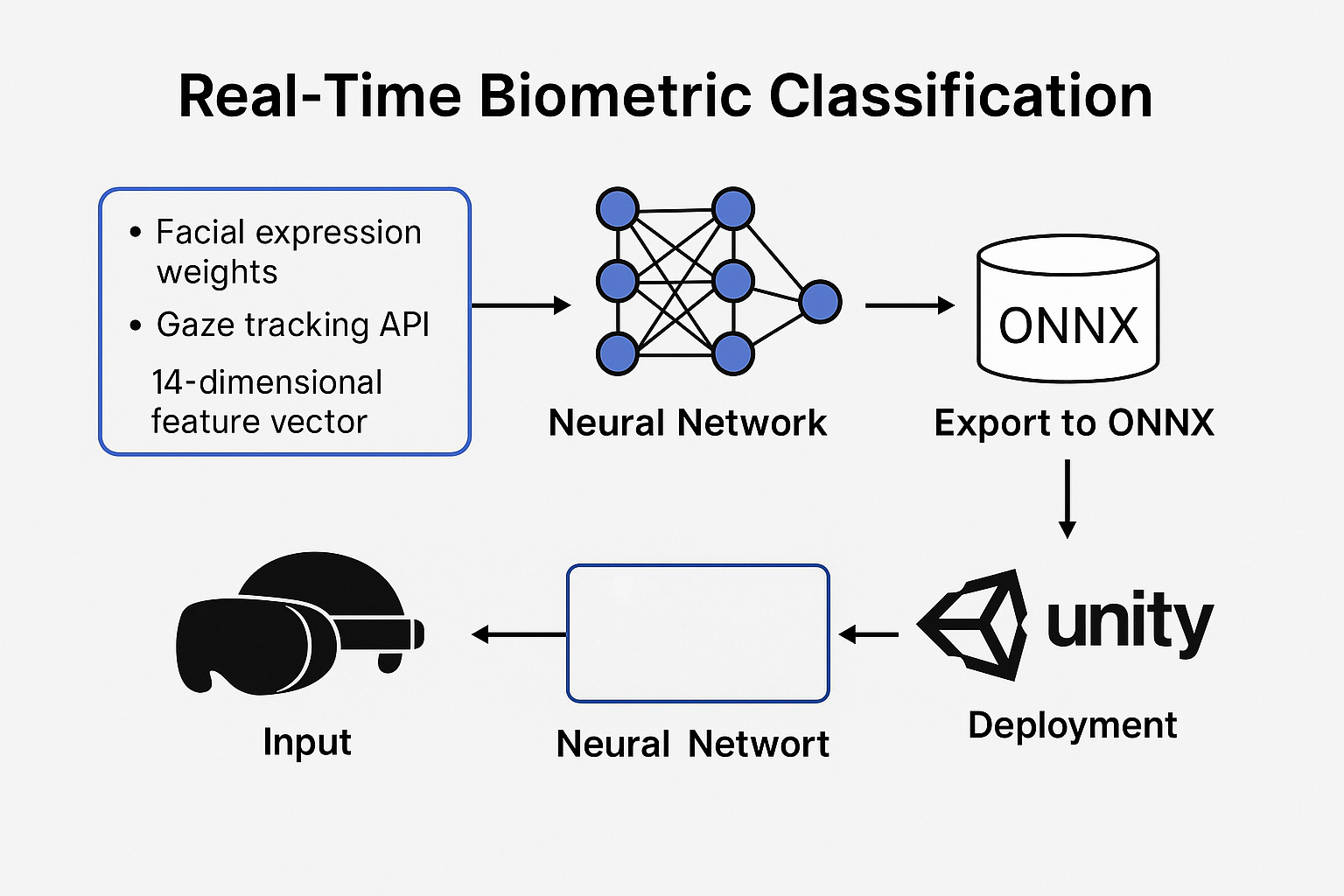}}
    \caption{Biometric Classification Flow.}
    \label{fig:placeholder}
\end{figure}

Once training was complete, the best-performing model was exported to Open Neural Network Exchange (ONNX) format using PyTorch’s \verb|torch.onnx.export()| utility. This intermediate format provides a hardware-agnostic, graph-optimized representation of the network, which is compatible with multiple inference backends. To integrate the model into our Unity-based VR application, we used Unity Sentis, a lightweight neural inference engine designed for on-device AI execution on standalone VR headsets like the Meta Quest Pro. The Sentis runtime loaded the ONNX model and performed inference directly on the headset's GPU/CPU without requiring internet connectivity or external SDKs. The end-to-end pipeline is as follows: 1) Facial and gaze vectors are collected in real-time via Unity's XR plugins. 2) The 14-dimensional vector is passed to the ONNX model using Sentis APIs. 3) The model outputs a probability distribution over user states (e.g., \verb|Neutral: 0.75|, \verb|Stressed: 0.10|, \verb|Engaged: 0.15|). 4) Based on the predicted class, the SDK determines the level of expression filtering to apply.\par

Evaluation and Feature Analysis: Model performance was evaluated using Accuracy, Macro-averaged F1-score, and the Confusion matrix analysis is conducted. The classifier achieved consistent accuracy across folds, with no significant variance across environment conditions, confirming generalizability.Further, feature importance analysis (using SHAP values and ablation testing) indicated that:

\begin{itemize}
   \item  Eye gaze direction (e.g., \verb|EyesLookDownR|, \verb|EyesLookRightL|) was a strong predictor of engagement and focus.
    \item Inner brow movement (e.g., \verb|InnerBrowRaiserL|) was highly correlated with stress and surprise.
    \item Cheek tension and lip pressor values were informative in distinguishing active from neutral states.
\end{itemize}
These findings validate our feature selection strategy and support the feasibility of using compact biometric vectors for real-time, privacy-sensitive classification. In Unity, the ONNX model was loaded and executed using Unity Sentis, shown in the following sub-section, which supports GPU and CPU inference on Android-based standalone headsets like the Meta Quest Pro. Real-time predictions were computed using \verb|InferenceEngine.Execute()| with live data streamed from the Meta XR SDK. The classifier was integrated into a Unity application that dynamically adapted behavior based on the predicted user state or filtered sensitive signals according to the experimental design.

\subsection{Unity Sentis Inference Code}

\begin{lstlisting}[language=C, caption={Unity C\# script using Sentis for ONNX inference}, label={lst:sentis_unity}]
using Unity.Sentis;
using UnityEngine;

public class BiometricClassifier : MonoBehaviour
{
    public NNModel onnxModelAsset;
    private Model runtimeModel;
    private IWorker worker;

    private float[] featureVector = new float[14];

    void Start()
    {
        runtimeModel = ModelLoader.Load(onnxModelAsset);
        worker = WorkerFactory.CreateWorker(BackendType.GPUCompute, runtimeModel);
    }

    void Update()
    {
        FillInputWithLiveData(); // Populate featureVector

        Tensor inputTensor = new Tensor(1, 14, featureVector);
        worker.Execute(inputTensor);

        Tensor output = worker.PeekOutput();
        int predictedClass = ArgMax(output.ToReadOnlyArray());

        Debug.Log($"Predicted biometric state: Class {predictedClass}");

        inputTensor.Dispose();
        output.Dispose();
    }

    int ArgMax(ReadOnlySpan<float> values)
    {
        float maxVal = float.MinValue;
        int maxIndex = 0;
        for (int i = 0; i < values.Length; i++)
        {
            if (values[i] > maxVal)
            {
                maxVal = values[i];
                maxIndex = i;
            }
        }
        return maxIndex;
    }

    void OnDestroy()
    {
        worker.Dispose();
    }
}
\end{lstlisting}


\subsection{Feature Summary and Interpretability }

The headset (Meta Quest Pro) continuously recorded facial expression weights mapped to standardized blendshape metrics. A single example from the dataset or say the feature vector is presented in the Table~\ref{tab:feature_vector}, we see distinctive emotional or cognitive signals moderate brow furrowing and inner brow elevation suggest concern or confusion, while slight lip and eye movements reflect attention shifts. The above Table~\ref{tab:feature_vector}  presents a sample feature vector representing quantitative expression values captured from facial muscle movements and ocular activity using a head-mounted device such as the Meta Quest Pro. Each feature corresponds to a specific facial action unit or eye movement, with values typically normalized between 0 and 1, indicating the relative intensity of the movement or activation at a given time frame.
For example, both \textit{Brow Lowerer Left (L)} and \textit{Brow Lowerer Right (R)} show moderate activation (0.26), suggesting a symmetrical lowering of the eyebrows. \textit{Cheek Raiser} values indicate slight asymmetry (0.08 on the left and 0.06 on the right), possibly reflecting a subtle emotional expression. \textit{Eyes LookRight L} shows a high value of 0.97, signifying a pronounced left eye gaze towards the right visual field. Meanwhile, the \textit{Eyes LookDown R} value of 0.31 indicates a concurrent downward motion of the right eye, suggesting complex gaze behavior. Other expressions, such as \textit{Jaw Drop} (0.02) and \textit{Lip Corner Depressor L} (0.02), show minimal activation, indicating a largely neutral lower facial region during this capture. Low values for features like \textit{Lip Suck LB} (0.01) and \textit{Tongue Tip Alveolar} (0.02) suggest no significant oral articulation at the time. 

\begin{table}[!t]
\captionsetup{font=small}
\caption{Sample Feature Vector Values for Facial Expressions. These feature vectors serve as input to real-time classification models for emotion detection, user intent inference, or biometric filtering. They also illustrate the richness and granularity of facial data that can be leveraged in immersive applications or filtered for privacy-preserving computation.}
\label{tab:feature_vector}
\begin{center}
	\resizebox{0.25\textwidth}{!}{%
		\begin{tabular}{c|c}
			\bfseries Feature & \bfseries Values \\
            \hline  
Brow Lowerer Left(L) & 0.26 \\
Brow Lowerer Right (R) & 0.26 \\
Cheek Raiser L & 0.08 \\
Cheek Raiser R & 0.06 \\
Eyes Closed L & 0.01 \\
Eyes Closed R & 0.08 \\
Eyes LookDown R & 0.31 \\
Eyes LookRight L & 0.97 \\
Inner BrowRaiser L & 0.40 \\
Jaw drop & 0.02 \\
Upper Lip Raiser R & 0.09 \\
Lip Corner Depressor L & 0.02 \\
Lip Suck LB & 0.01 \\
Tongue Tip Alveolar & 0.02 \\
\end{tabular}%
	}
	\end{center}
\end{table}

\subsection{Psychological Interpretations}
\begin{figure}[htbp]
    \centering
    \includegraphics[width=3.2in, height=1.8in]{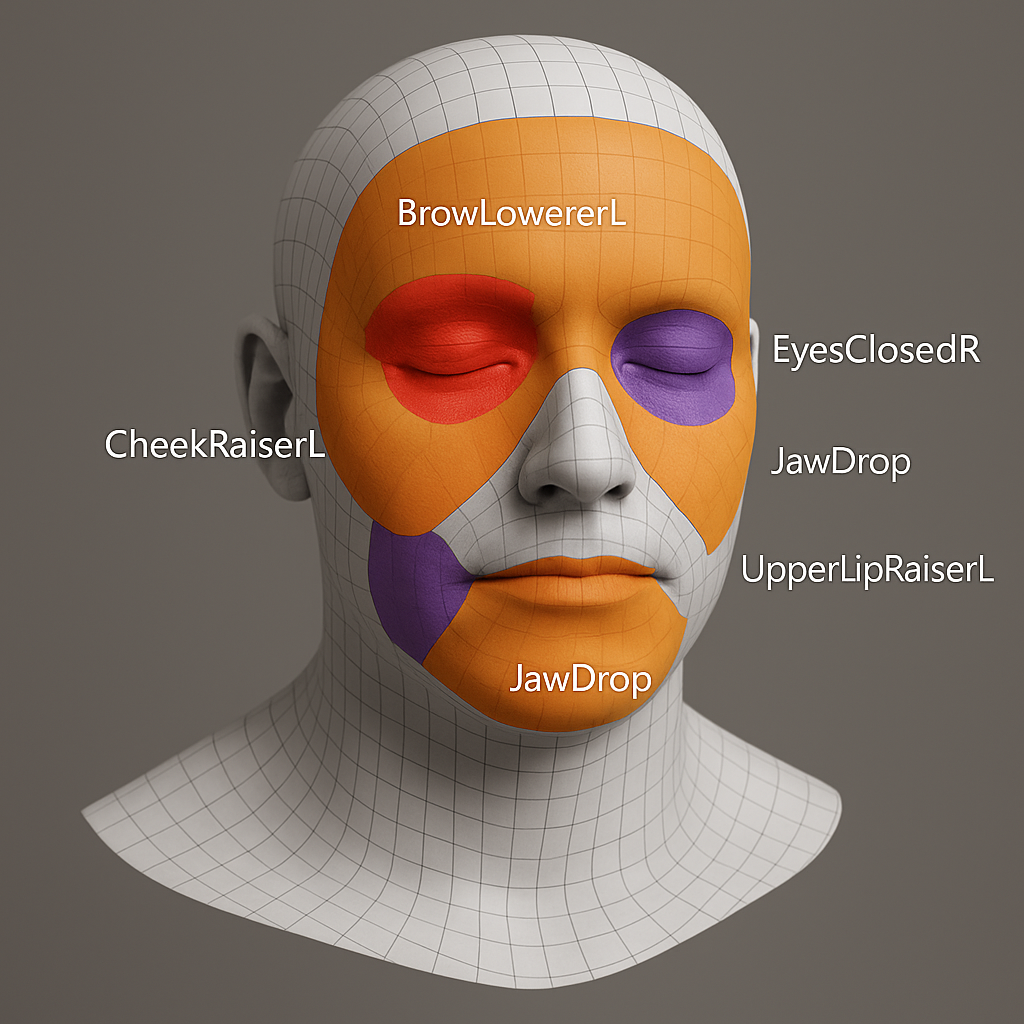} 
    \caption{Annotated 3D visualization of facial expression weights captured from a representative XR session. Highlighted muscle regions correspond to actively engaged facial features, with associated intensity values (range: 0–1) indicating the degree of activation. This snapshot reveals moderate brow tension, subtle cheek and lip movement, and distinct gaze orientation demonstrating how facial data can encode cognitive and emotional cues relevant to biometric inference.}
    \label{fig:facial_map}
\end{figure}

We compiled the collected features into an interpret-ability mapping, combining physiological descriptions with possible emotional or psychological inferences. The  Table~\ref{tab:psych_cues} presents a condensed mapping between facial expression features captured by the XR headset and their corresponding physiological and psychological interpretations. 
\begin{table}[!t]
\captionsetup{font=small}
\caption{Facial Features with Physiological Meanings and Psychological Cues}
\label{tab:psych_cues}

\begin{center}
	\resizebox{0.47\textwidth}{!}{%
		\begin{tabular}{c|c|c}
			\bfseries Feature & \bfseries Physiological Meaning & \bfseries Psychological Cue \\
            \hline  
            \centering

BrowLowerer & Brow tension & Confusion, focus \\
InnerBrowRaiser & Inner brow lift & Sadness, surprise \\
CheekRaiser & Cheek elevation & Genuine smile, joy \\
EyesClosed & Blinking & Fatigue, disengagement \\
EyesLook Direction & Gaze movement & Attention, curiosity \\
JawDrop & Mouth open & Shock, speech readiness \\
LipCornerPuller & Smile tension & Positivity, friendliness \\
LipPucker & Lips pushed forward & Doubt, contemplation \\
LidTightener & Squint & Irritation, tension \\
UpperLipRaiser & Upper lip pull & Disgust, disapproval \\
Tongue Tip Motion & Tongue articulation & Speaking intent, discomfort \\
\end{tabular}%
	}
	\end{center}
\end{table}
Further, the annotated 3D visualization of facial expression weights
captured from a representative XR session is shown in Fig.~\ref{fig:facial_map}. Beyond single-feature interpretation, we evaluated combinations of signal patterns that tend to correspond with complex states: 
\begin{enumerate}
    \item Inner Brow Raiser + Cheek Raiser + Eyes Closed → Emotional reaction (e.g., empathy, sadness)
   \item Lip Pucker + Jaw Drop + Lid Tightener  → Hesitation or decision pressure
    \item Eyes Look Right + Jaw Drop + Tongue Tip Alveolar  → Cognitive engagement or speech preparation 
\end{enumerate}

Previous research has demonstrated that facial expression data, whether in static images, video streams, or interactive applications, can be used to infer sensitive attributes such as emotional state, gender, mental health indicators, and behavioral traits~\cite{wang2023identifying, wen2024face, deng2023face, singh2025facial}. \textit{While many of these studies were conducted outside XR contexts, their findings underscore the broader risks associated with collecting rich facial signals. When such data are continuously captured in immersive environments, these risks can be amplified, making it essential to treat facial expressions as high-sensitivity biometric data and enforce privacy-aware mechanisms before transmission or storage}~\cite{sood2025framework}. 

\begin{figure*}[h]
\centering
\subfloat[]{\includegraphics[width=3.35in, height=2.2in]{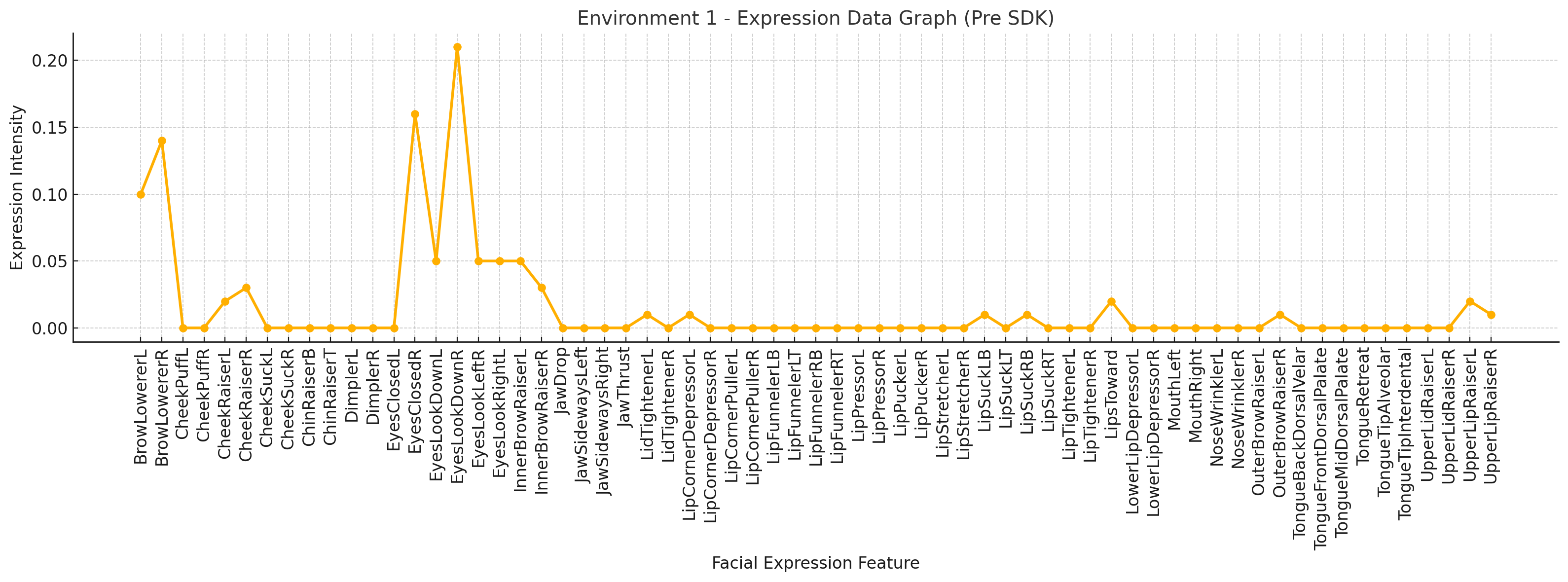}}
 \subfloat[]{\includegraphics[width=3.35in, height=2.2in]{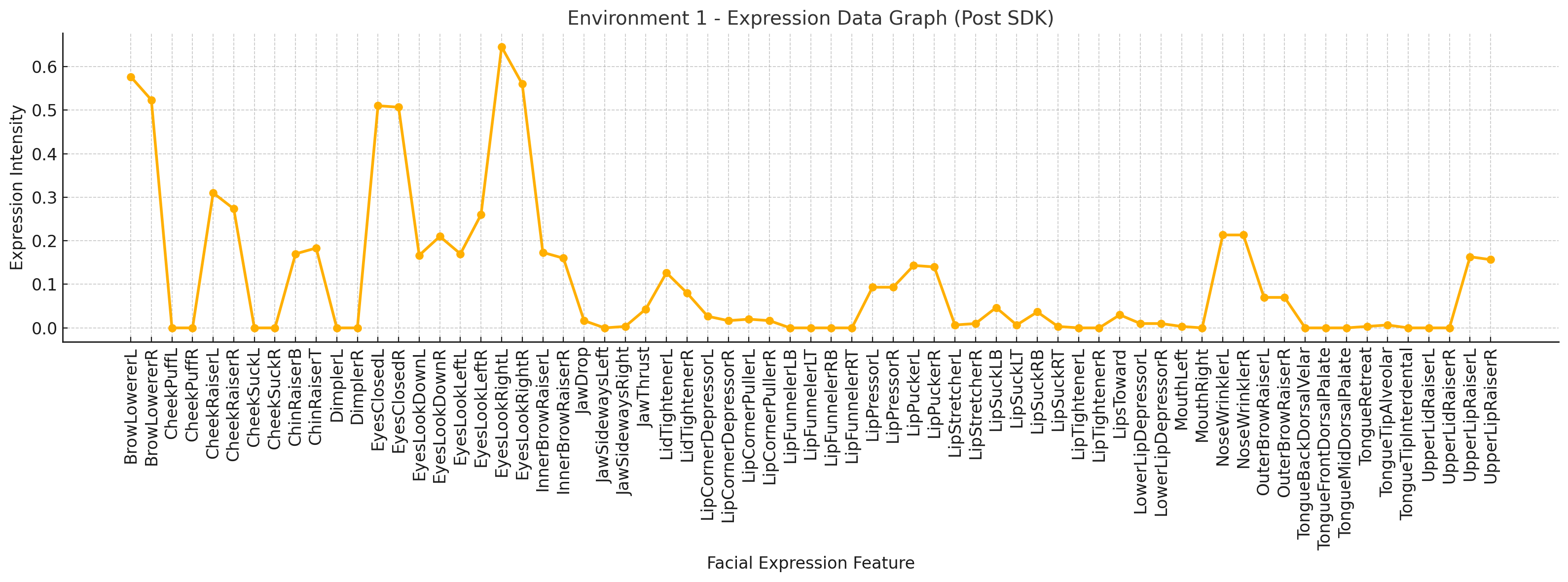}}
\caption{a) Environment 1 – Expression Graph (Pre SDK), b) Environment 1 – Expression Graph (Post SDK)  }
\label{fig:enter-label1}
\end{figure*}

\begin{figure*}[h]
\centering
\subfloat[]{\includegraphics[width=3.35in, height=2.2in]{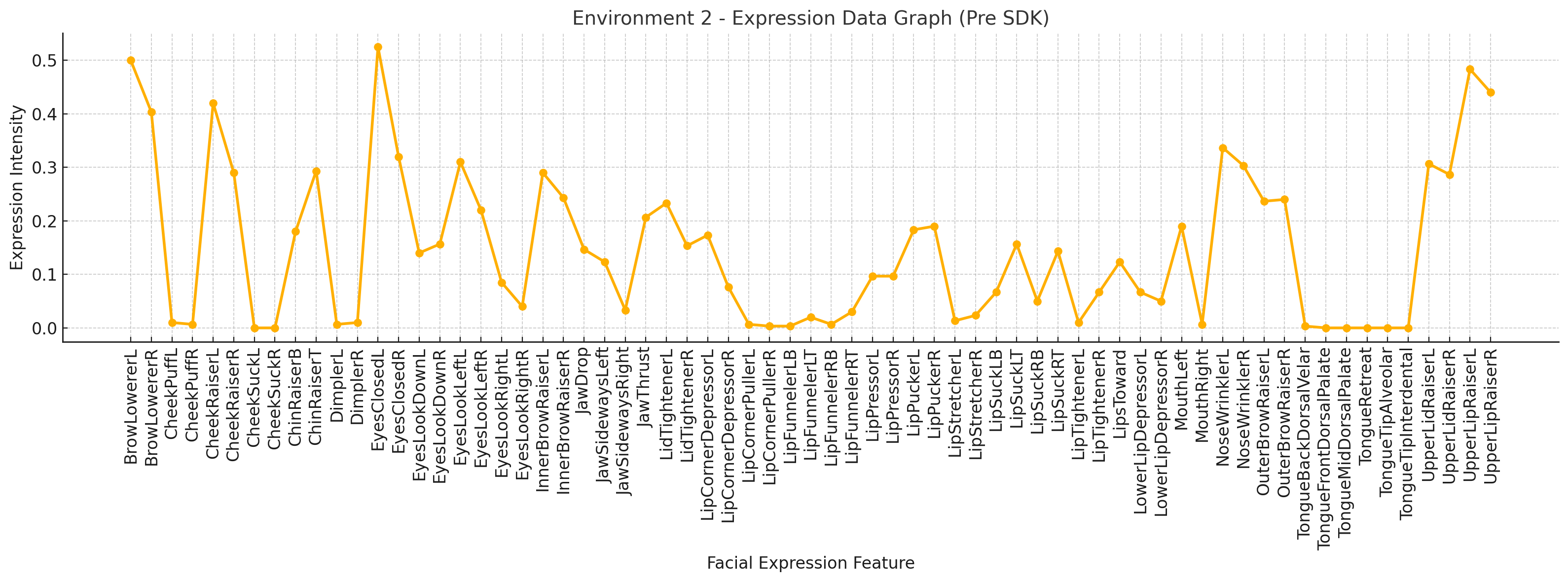}}
 \subfloat[]{\includegraphics[width=3.35in, height=2.2in]{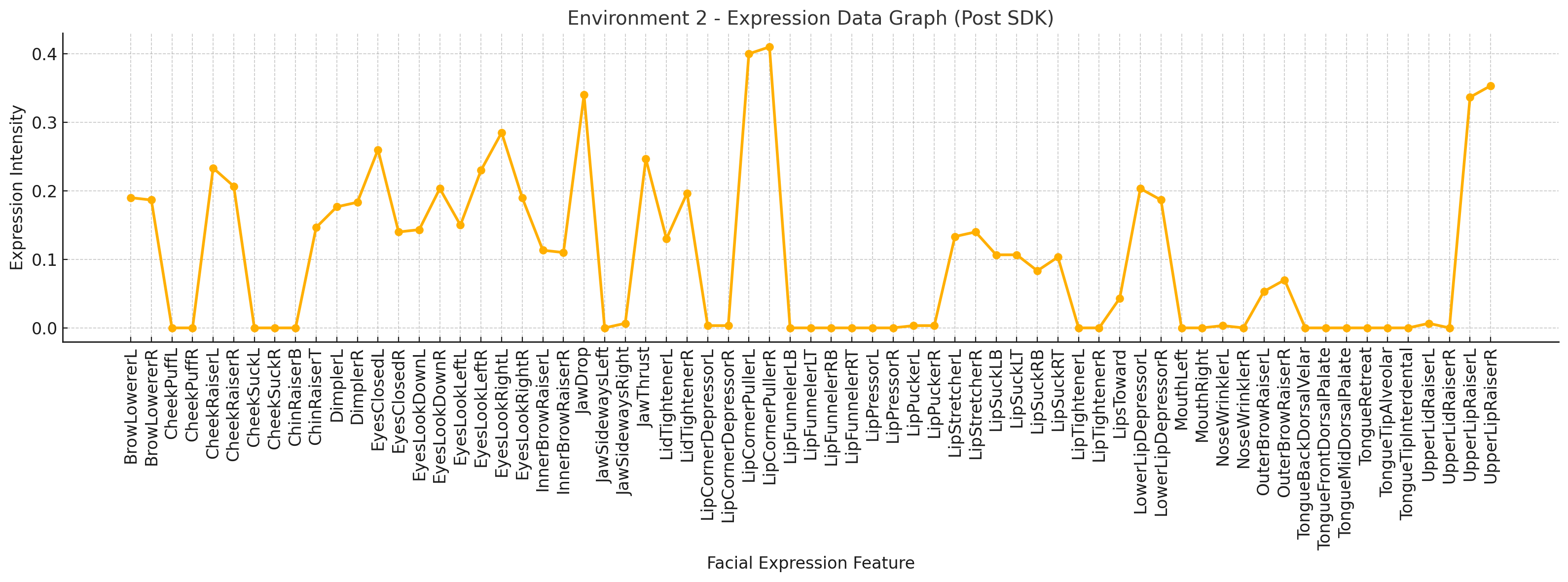}}
\caption{a) Environment 2 – Expression Graph (Pre SDK), b) Environment 2 – Expression Graph (Post SDK)  }
\label{fig:enter-label2}
\end{figure*}

\begin{figure*}[h]
\centering
\subfloat[]{\includegraphics[width=3.35in, height=2.3in]{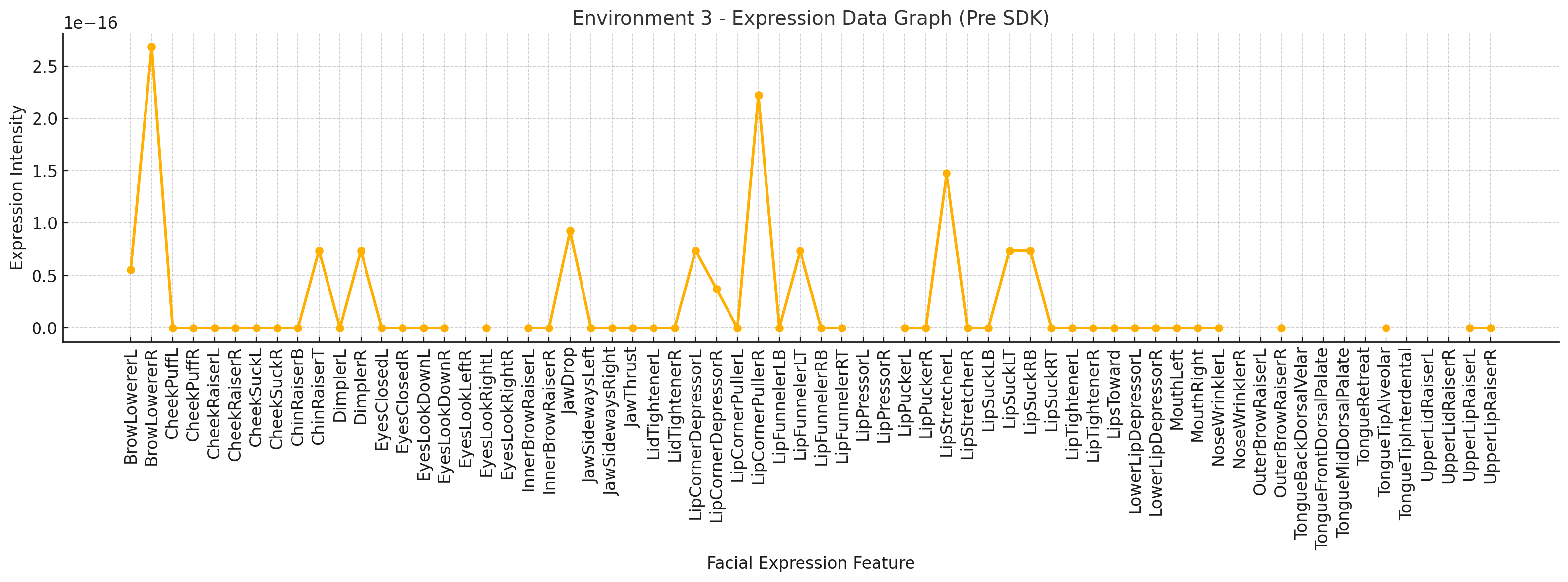}}
 \subfloat[]{\includegraphics[width=3.35in, height=2.3in]{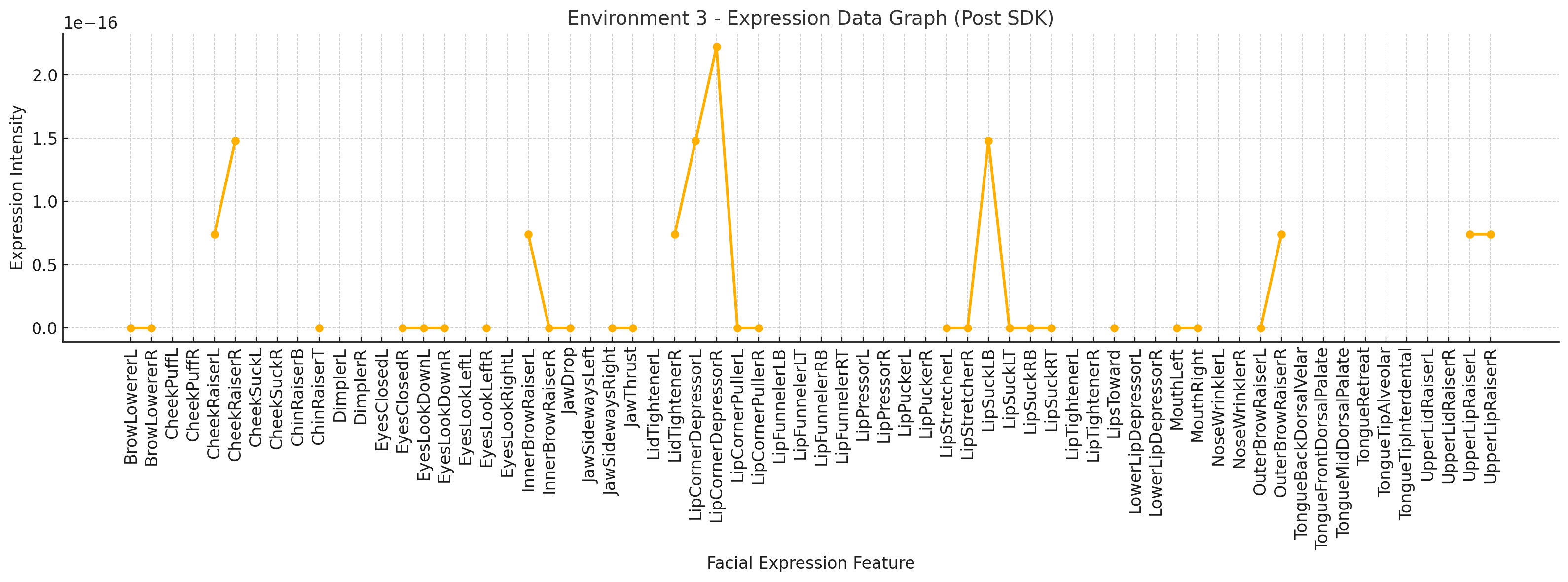}}
\caption{a) Environment 3 – Expression Graph (Pre SDK), b) Environment 3 – Expression Graph (Post SDK)  }
\label{fig:enter-label3}
\end{figure*}

\section{Performance Evaluation and Results}
To assess the efficiency and real-time viability of our proposed privacy-preserving biometric filtering SDK, we have conducted a comparative analysis across three immersive virtual environments, each designed to trigger distinct emotional and cognitive responses. These environments included: (1) a task-oriented interactive simulation, (2) a social engagement scenario, and (3) a meditative, low-stimulus setting. For each scenario, facial expression data was recorded under two controlled conditions: (1) Pre-SDK, representing unfiltered raw biometric signals, and (2) Post-SDK, wherein the SDK’s real-time privacy-preserving filtering was actively running on the edge device before any data left the headset. This setup enabled a fine-grained comparison between unfiltered and privacy-protected biometric streams, allowing us to analyze signal consistency, system performance, user interaction quality, and emotional expressivity. 

\subsection{Expression Signal Consistency Across Conditions }
Now, we show the ``Expression Signal Consistency Across Conditions". As seen in Figure~\ref{fig:enter-label1},~\ref{fig:enter-label2}, and~\ref{fig:enter-label3} we show that each figure depicts facial expression intensity values derived from aggregated expression weight data. Each line graph represents the mean activation intensity for 50+ expression vectors, which span facial features ocular, oral, and muscular. The biometric filtering mechanism being active during the post-SDK sessions, expression patterns remain remarkably consistent across pre and post-conditions. Notable peaks and troughs in facial feature activations, such as brow lowering, eye closure, cheek raising, or lip movement, persist in both data streams. This consistency indicates that the SDK filtering pipeline retains the naturalistic rhythm and amplitude of user expressions, preserving essential user-environment feedback loops. These figures visually affirm that the application experience remains stable regardless of whether raw expression data is streamed or suppressed locally. \par
\begin{figure}
    \centering
     \includegraphics[width=3.2in, height=2.4in]{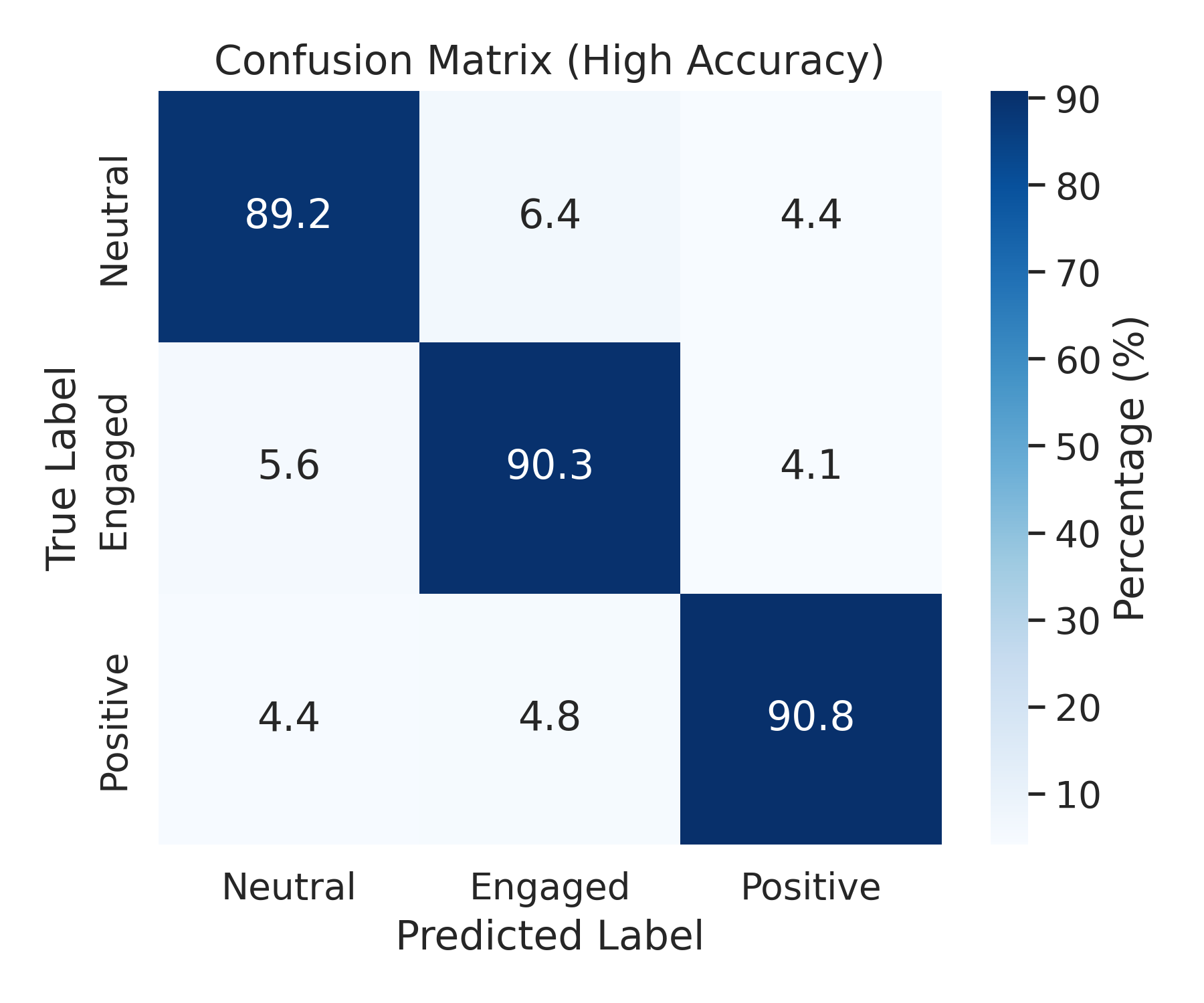}
    \caption{Confusion matrix illustrating the classification performance of the proposed on-device biometric filtering framework across three affective states: Neutral, Engaged, and Positive. Each cell represents the percentage of samples in a given true class (rows) that were assigned to a predicted class (columns). The strong diagonal dominance indicates high classification accuracy and minimal confusion between categories, confirming that the privacy-preserving SDK maintains model discriminative power while filtering sensitive biometric signals in real time. }
    \label{fig:8}
\end{figure}

 
\textit{Preserving Immersive Feedback and Emotional Expression:} Now we also show that the framework is efficient in preserving Immersive Feedback and Emotional Expression. In all three environments, participants were able to perform context-appropriate actions without degradation in performance, responsiveness, or perceived realism. Even in Environment 3, which simulated a meditative setting with minimal stimuli, participants displayed neutral yet expressive biometric profiles in both pre- and post-SDK states. For example, the relative flatness of the intensity values in Figures~\ref{fig:enter-label3} (a) and (b) reflect the calming, minimal expression nature of the scenario that validates that SDK filtering does not inject noise or artificially suppress expected neutral behavior. In more dynamic contexts (e.g., Environment 1 and 2), which required task execution or reactive gestures, users consistently engaged with their surroundings, and their affective facial cues remained detectable even with the SDK operating in filtering mode. Fig.~\ref{fig:8} shows the confusion metrics of the evaluation of our approach.  \par 
\textit{Absence of Functional Degradation:} Crucially, no functional disparity was observed in application behavior or user interaction. Applications that relied on real-time expression feedback (e.g., gaze tracking, avatar mirroring, or UI interaction) responded identically in both testing conditions. This confirms that the SDK's filtering approach does not interfere with core performance features or with the perceptual realism of avatars or environments. The SDK's modular implementation ensures that only personally identifiable biometric signatures are removed, without discarding the general-purpose signal structure required for runtime logic. Thus, our results confirm that it is feasible to integrate privacy-sensitive filtering without requiring application-level modifications or compromising interaction quality. 

{\textit{Why Developers Need This Framework }:} From a developer’s perspective, integrating biometric inputs into immersive environments unlocks powerful new interaction models  from emotion-driven storytelling to adaptive avatars. However, these very inputs also pose significant privacy and compliance risks, especially as biometric data is increasingly regulated under laws such as the GDPR, CCPA, and Australia’s Privacy Act 1988. Our SDK addresses this gap by enabling developers to 1) build trust with users by providing transparent, local privacy enforcement, 2) ensure compliance with global biometric data regulations without redesigning application logic, 3) safely deploy emotion-aware features, (e.g., avatar mirroring, wellness tracking) without collecting or transmitting identifiable biometric signatures, and 4) avoid cloud-based processing bottlenecks by filtering on-device, thus preserving bandwidth and reducing latency. Moreover, the SDK is implemented as a modular Unity-compatible plug-in, requiring no modification to the application’s core expression logic. This means developers can immediately benefit from privacy safeguards without rewriting facial animation systems, input pipelines, or UI behaviors  making it a future-proof, production-ready addition to any Unity XR development workflow. We are addressing privacy concerns in the early stages of application development and system design.

\begin{table*}[!t]
\captionsetup{font=small}
\caption{Comparison of Related Work and Our Framework}
\label{tab:comparison}

\begin{center}
\resizebox{\textwidth}{!}{%
\begin{tabular}{c|c|c|c|c}
\bfseries Aspect & \bfseries Wu et al.~\cite{wu2023privacy} & \bfseries FaceReader~\cite{zhang2023facereader}& \bfseries IMWUT~\cite{o2023privacy}& \bfseries Our Work\\
\hline
System Type & Adversarial demo & Emotion analytics tool & Social perception study & Privacy-preserving SDK \\
Data Source & Motion and controller data & Facial landmarks & Interviews & Expression weights + gaze \\
ML Inference & Unsupervised learning & Time-series regression & None & Lightweight DNN (ONNX) \\
Deployment Mode & Middleware-level attack & Offline post-processing & Not applicable & On-device via Unity Sentis \\
\end{tabular}%
}
\end{center}
\end{table*}

\section{Comparison with Prior Work and Novelty of Our Approach}
Biometric data privacy in immersive environments has received growing attention in recent years, as researchers uncover how sensor-rich XR devices expose users to previously unanticipated risks. While several prior studies have explored the extraction and interpretation of biometric signals from AR/VR systems, most existing implementations either serve as proof-of-concept attacks or retrospective analytics pipelines. A high-level comparison is given in Table~\ref{tab:comparison} and detailed discussion is given below. In contrast, our work is focused on building an operational, privacy-aware SDK that functions in real time, operates entirely on-device, and integrates directly with Unity-based XR applications. 

\subsection{Biometric Inference and Threat Demonstration }

The authors~\cite{wu2023privacy} presents the most prominent examples of motion sensor-based side-channel inference in VR. The authors show that unrestricted access to motion and controller data (via OpenVR and Oculus SDKs) can be exploited to reconstruct keystrokes typed on virtual keyboards. Their implementation is notable for achieving high accuracy (up to 89.7\%) using an unsupervised learning approach, without any user-specific training. However, the work is adversarial in nature: the researchers intentionally mimic attacker behavior to reveal privacy vulnerabilities but stop short of offering real-time or embedded mitigations. Their architecture operates at the middleware level and is designed to capture unrestricted sensor telemetry across platforms, making it powerful but unsuitable for legitimate deployment in privacy-conscious applications. \par
In contrast, our approach is defensive. Rather than trying to infer private signals, our SDK aims to prevent their transmission in the first place by classifying biometric states on the device and applying adaptive filtering before data exit the headset. While~\cite{wu2023privacy} demonstrate the feasibility of extracting identity or intent from biometric signals, our work addresses the next logical step in the privacy pipeline namely, real-time suppression of sensitive biometric features using machine learning.

\subsection{Emotion and Vital Sign Recognition in AR/VR }

The FaceReader~\cite{zhang2023facereader} showcases the use of facial motion data to unobtrusively extract emotional and physiological signals such as heart rate variability. It relies on high-resolution facial landmark tracking, which is analyzed over time to derive sentiment and physical metrics. While the authors demonstrate high fidelity in signal recovery, their system is intended for offline use. Facial data is logged during runtime and then processed externally using batch analysis pipelines. No classifier inference is performed on-device or in real time.\par
Our implementation differs in both intent and infrastructure. First, rather than mining or maximizing signal yield, we are deliberately classifying expression vectors for the purpose of selective suppression. This aligns with privacy-preserving design principles where only non-sensitive or generalized biometric signals are permitted to persist. Second, our use of Unity Sentis enables the model to run directly on the Meta Quest Pro headset, requiring no server offloading, backend processing, or cloud inference. This decentralization is particularly important in healthcare, education, and safety-sensitive VR applications, where compliance with data sovereignty and user consent laws is critical. Furthermore, the FaceReader model operates at a level of granularity that is incompatible with real-time immersive rendering. Our lightweight 14-dimensional input model is explicitly chosen to prioritize computational efficiency, thereby preserving frame rates and input latency in dynamic XR environments.

\subsection{Social Acceptability Studies in AR Devices }

In a separate thread of research, the IMWUT AR Acceptability study~\cite{o2023privacy} takes a qualitative approach to understanding user concerns about wearable AR devices. The authors explore factors such as context sensitivity, perceived social intrusiveness, and trust in device providers. Although this work provides essential information on the societal framework of biometric privacy, it does not propose or evaluate a technical solution to mitigate biometric risk. Instead, it supports the idea that privacy should be embedded into system design, a principle our framework operationalizes through real-time biometric classification and filtering.

Our SDK can be seen as a direct response to these concerns, offering XR developers a practical toolkit to implement expression-aware interaction without exposing raw facial vectors to potential misuse. Importantly, the modularity of our implementation allows privacy filters to be toggled, tuned, or extended without altering application-level code, enabling compliance with future regulatory frameworks such as ISO/IEC 27701 and GDPR Article 25 on privacy by design.

 \section{Discussion: Challenges and Opportunities}
In this section, we highlight the limitations of the current work which serve as our objectives to continue this work in the future and this will encourage research community to contribute in this very direction. \par 
1) The current implementation is tightly coupled with Unity and the Meta SDK. Although this provides strong compatibility and performance for a subset of XR applications, it imposes limitations on generalizability and wider industry adoption. We plan to extend SDK compatibility to support additional platforms and engines, including Unreal Engine, OpenXR, and WebXR. This would allow developers across a wider spectrum of XR tools and ecosystems to adopt privacy-preserving practices without the need to re-architect their applications. Another area of future research lies in the enhancement of the breadth and depth of the biometric classifier. Currently, the classifier focuses on facial expression weights and eye tracking metrics. We aim to expand this to include additional physiological signals such as heart rate variability, galvanic skin response (GSR), and respiration patterns, especially as emerging XR headsets begin to include these biosensors as standard. Incorporating these modalities would allow more comprehensive protection and allow the framework to better adapt to the evolving sensor landscape in XR.\par
2) To improve classifier robustness and fairness, we also intend to expand the training data set with more diverse user profiles, including individuals with expression patterns specific to atypical or disabilities. This step is essential to avoid model bias and ensure equitable protection for users of all backgrounds and abilities. In addition, leveraging continuous learning techniques can help the system adapt to user-specific variations over time without compromising privacy. In parallel with technical improvements, we recognize the importance of aligning the SDK with regulatory and industry standards.\par
3) Further, we envision strategic collaborations with XR hardware vendors to enable low-level enforcement of biometric filtering mechanisms potentially at the firmware or driver level. Native enforcement would significantly reduce attack surfaces and offer hardware-level guarantees, moving beyond application-layer protection. \par
4) Nonetheless, the technical intervention (as we have discussed in the paper) is extremely challenging (if not impossible) because a) the autonomous data labeling is challenging in this dynamic VR environment and b) on the basis of the volume of sensors and massive volumes of data the VR devices collect as well as the variability of each human subject with swarms of data and the risk of algorithms throwing outliers under the bus. \par
In-spite of the above gaps, our work is an early work and we encourage researchers to collaborate and investigate solutions from different aspects. In summary, our work demonstrates the feasibility and importance of real-time biometric filtering in XR by proactively addressing privacy concerns in the early stages of application development and system design.

\section{Conclusion and Future Work}

In this work, we proposed a novel privacy-preserving biometric data filtering system designed specifically for immersive extended reality applications. As XR systems continue to integrate increasingly sophisticated sensor arrays such as facial tracking, eye tracking, and physiological monitors, they also introduce new privacy challenges due to the richness and sensitivity of the data being collected. Our solution addresses the growing threat of biometric data leakage by introducing a modular framework that operates in real time within Unity-based applications, using the Meta Quest Pro as a development and testing platform. At the core of our framework is a lightweight machine learning classifier that identifies biometric signals including facial expressions and ocular metrics with high accuracy. Based on classification outcomes, the system selectively filters or suppresses sensitive signal components before they are transmitted or accessed by downstream applications. This ensures that users retain granular control over the aspects of their biometric profile shared, in accordance with the principles of privacy-by-design. We have highlighted the future work in further discussion section which we aim to do to continue this work. \par

\vskip -2\baselineskip plus -1fil
\ifCLASSOPTIONcaptionsoff
  \newpage
\fi
\bibliographystyle{IEEEtran}
\bibliography{references.bib}
 \vskip -2\baselineskip plus -1fil
\begin{IEEEbiography}[{\includegraphics[width=1in,height=1.25in, clip]{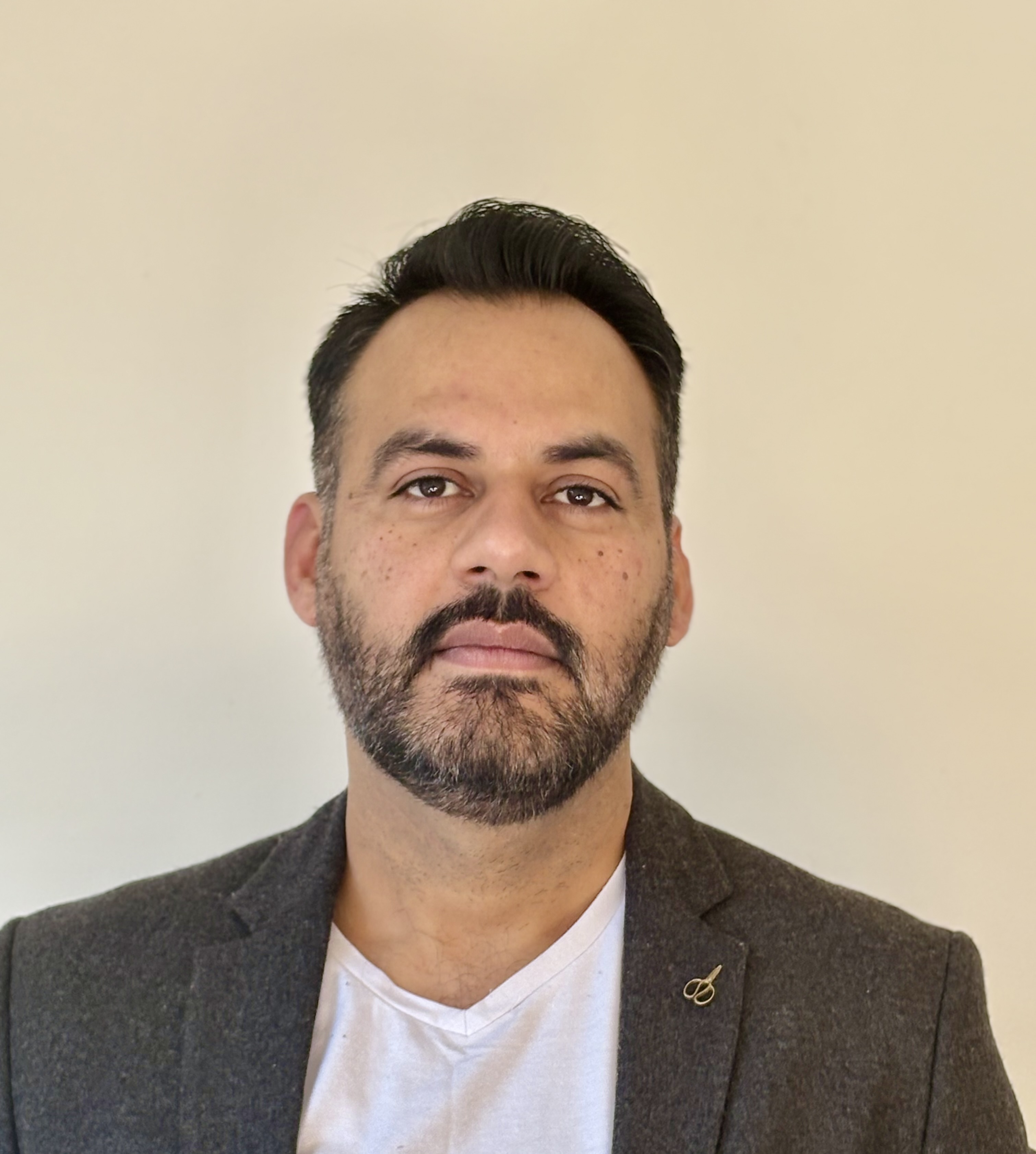}}]{Keshav Sood} (Senior Member IEEE) received his B.Tech. degree (Hons.) in Electronic and Communication Engineering with distinction and the M.Tech. degree in Optical Fiber Engineering in 2007 and 2012, respectively. He was a trainee with the Terminal Ballistic Research Laboratory (TBRL, DRDO, Ministry of Defense) in Chandigarh, India. He received his Ph.D. degree in Information Technology (software-defined networking security) from Deakin University, Melbourne in 2018. He was the recipient of the Professor of IT award given by the School of IT for his outstanding academic achievements during his PhD degree. He completed his post-doctoral from The University of Newcastle, New South Wales, Australia. He is currently working as a Senior Lecturer in Cyber Security at Deakin University, School of IT. His work in cyber security for next generation networks has been published in top-notch security and networking venues.
\end{IEEEbiography}
\vskip -2\baselineskip plus -1fil
\begin{IEEEbiography}[{\includegraphics[width=1in,height=1.25in, clip]{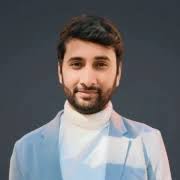}}]{Sanjay Selvaraj} holds a Bachelor’s degree in Computer Science from Dr Mahalingam College of Engineering and Technology, India with a specialization in 3D Modeling in 2020, and a Master’s degree in Information Technology from Deakin University, where he specialized in Virtual Reality systems and applications in 2023. Currently he is working as a research assistant,  in VR and immersive development, at Deakin University, Melbourne. He is a VR developer specializing in the creation of immersive experiences that challenge the boundaries of digital reality. His recent projects explore safety education in the metaverse, adaptive storytelling, and real-time emotion-aware avatars.

\end{IEEEbiography}
\vskip -2\baselineskip plus -1fil
\begin{IEEEbiography}[{\includegraphics[width=1in,height=1.25in, clip]{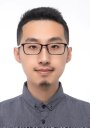}}]{Youyang Qu} received the BEng degree in mechanical automation in 2012, the MEng degree in software engineering in 2015, both from Beijing Institute of Technology, China, and the PhD degree from Deakin University, Australia in 2019. He is currently a research scientist at Data61, Commonwealth Scientific and Industrial Research Organization (CSIRO), Australia. Before joining CSIRO, he served as a research fellow at Deakin University, Australia. His research interests focus on machine learning, big data, IoT, blockchain, and corresponding security and customizable privacy issues. He has over 50 publications, including high-quality journals and conferences papers. He is active in the research society and has served as an organizing committee member in SPDE 2020, BigSecuirty 2021, and Tridentcom 2021/2022.
\end{IEEEbiography}
\end{document}